# REAL-TIME ENERGY MONITORING INFRASTRUCTURE FOR RESIDENTIAL COLLECTIVE SELF-CONSUMPTION OPERATIONS USING LINKY METER

*Jérôme Ferrari[1], Benoit Delinchant[1], Frédéric Wurtz[1], Olga Rouchouze[1]*

[1]Univ. Grenoble Alpes, CNRS, Grenoble INP, G2Elab, F-38000 Grenoble, France

## Abstract

As part of the energy transition and the rise in energy prices, the number of collective self-consumption operations in France is steadily increasing. However, energy flow monitoring currently relies on historical "day+1" data provided by Linky meters, which does not offer real time feedback to help participants adapt their energy consumption behaviors. This article introduces a new open-source infrastructure for real-time monitoring based on Linky meter data, enabling participants to make informed decisions and take timely actions. It includes a description of the xKy device, applied to a collective self-consumption operation involving nine participants, supported by the Energy Transition Observatory (OTE). The project encompasses the implementation of gateways in participants' homes and the development and operation of real-time monitoring website, aimed at increasing participants' self-consumption rate.

## 1 Introduction

Collective self-consumption is a method of economic valuation that consists of managing production distribution from a purely monetary point of view between producers and consumers who are physically close (less than 2 km).

In recent years, with the increase in electricity prices and based on data provided by the French Distribution System Operator (DSO) Enedis [1], the number of collective self-consumption operations has continued to increase in France since their appearance in 2018, rising from around 50 in March 2021 to 698 in December 2024 Fig. 1.

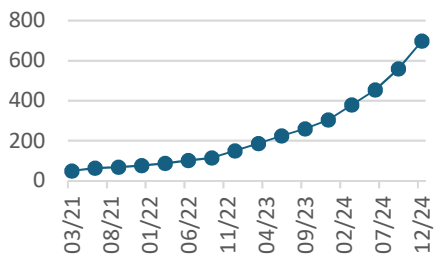

Fig. 1 collective self-consumption operations number

The emergence of this new type of operation was made possible by the installation by Enedis of communicating meters called "Linky" whose implementation began in 2015. Linky, through its communicating aspect, allows the consumption and production indexes of the past day to be sent to the DSO.

*1.1 Structure of a collective self-consumption operation*

Legally, an operation has 3 separate entities:
- The Organizing Legal Entity (OLE) which represents producers and consumers
- The Distribution System Operator which is counting the exchanges of electricity flows
- The energy suppliers who will send the invoices to consumers based on transmitted indexes.

At the technical level, it starts at Linky meters which send to the DSO servers every evening between midnight and 1 am the various daily production and consumption indexes of network users with a 10-minutes time step.

The indexes collected are then transferred once a month to the OLE. Based on the transmitted indexes, the OLE decides on the distribution keys to apply and transmit it to the DSO. Once the distribution keys have been obtained, the DSO transmits the corrected indexes to the various energy suppliers. Once these steps have been completed, the producers and energy suppliers send their invoices to the consumers Fig. 2.

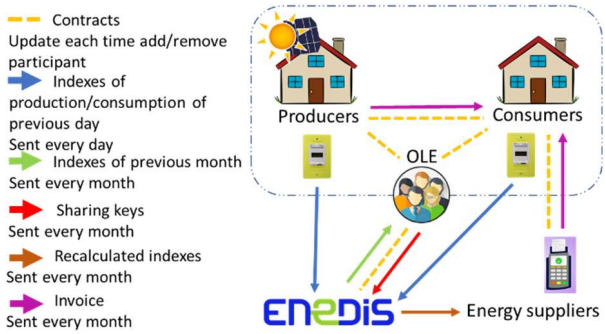

Fig. 2 Structure of the interactions of the collective self-consumption operation

*1.2 Current status of data usage in collective self-consumption operation*

*1.2.1 Procedure for setting up a collective self-consumption operation:* In the field of collective self-consumption of the residential type, when setting up a project, the actors mainly choose an optimization strategy based on a distribution of the number of producers/consumers as well as on the optimization of the sharing keys.

In the user guides provided by Enedis [2] and by the PEGASUS project (Microgrids in rural or island environments) [3], the guidelines for carrying out a collective self-consumption project are described in 4 consecutive phases which are as follows:
- A first phase of characterizing the project where the different actors are identified as well as the organizational diagram
- A second phase of feasibility study, to know the technical constraints, to characterize consumption profiles, to carry out the dimensioning of the production and to set up the economic model (Sales diagram, choice of the distribution key, definition of the producer/consumer perimeter)
- A third phase of realization, including the production set up, the creation of the OLE, the requests for access to the data to Enedis and the signing of the various contracts.
- A final phase of exploitation, where the daily data is used at each end of the month to create the distribution keys allowing the invoicing of producers and energy suppliers to consumers.

*1.2.2 Current data use in collective self-consumption:* The sole use of data from DSO only allows optimization through the calculation of the sharing keys once the exploitation phase has started.
One of the challenges is to maximize the self-consumption rate. The use of this data provided retrospectively and with a sampling interval of 30 min (15 min since the 9th of October) implies a "blind spot" in optimization.

Indeed, no action (direct or indirect) was made on the loads to follow production, nor for a detailed analysis of consumption to find optimization levers.
As stated by Nana Kofi's PhD thesis [4], it can be assumed that real-time monitoring (collectively and individually) should make the self-consumption rate increase but also should result in individual sufficiency increase.

## 2 Methodology

*2.1 Structure of the experiment xKy*

In the first xKy experiment [5], to overcome a "blind spot" of data already existing in the field of electrical energy, G2Elab developed an ERL gateway (Linky Radio Transmitter) called Winky based on the other capacity of the Linky meter to be able to provide various data in real-time (index, apparent power, tariff option, etc.) through a specific socket called TIC. The option of an ERL was chosen because its installation does not require the intervention of an electrician and can be carried out directly by the participants of the experiment.
This ERL as well as the data processing chain were developed with an "open-source" and "open-hardware" vision. This Open approach must allow participants in future experiments to know where and how their data is processed.
The ability to communicate with connected objects through the MQTT protocol was added to allow participants who potentially have a house with controllable devices to be able to control in real-time.
A first test of the structure began in the format of a participatory science experiment in October 2023 with 250 makers or technophiles [6] [7] and made it possible to create a first structure Fig. 3.

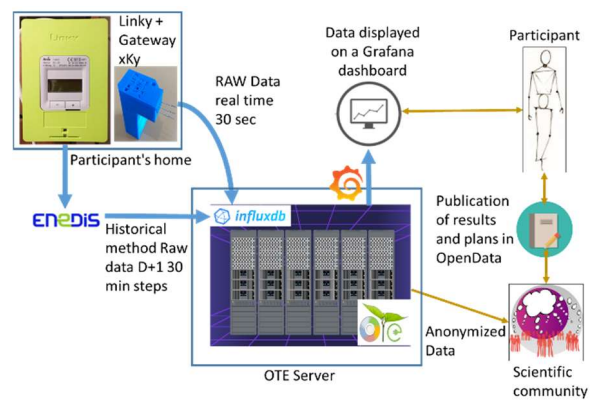

Fig. 3 Structure of the xKy experiment

A first open dataset resulting from this experiment was published on "research data gouv" repository website [8].

*2.2 Application of the xKy structure to the collective self-consumption experiment.*

The case study of collective self-consumption presented in this article has the following distribution:
- 6 producer houses, 1 of which has a storage battery
- 3 purely consumer houses

This distribution is interesting because it covers all the use cases offered by the Linky meter.

*2.2.1 Installation of device:* Unlike the previous experiment where the participants were driving forces and knew IOTs, the collective self-consumption experiment of this study also involved novices and required support for the implementation of the device among the various participants.

The process of setting up the device goes through 5 main steps which are as follows:
- Registration on the OTE website with an initial global questionnaire
- Completion of the xKy study questionnaire in order to receive login details
- Configuration of the gateway to connect to the participant's Wifi
- Installation of the gateway on the Linky
- Connection to the Grafana website to access the visualization interfaces

*2.2.2 Interface design:* For this operation, 2 types of interfaces designed under Grafana are made available to participants to allow them to follow their installation in real-time as well as to have an overall view of the operation

The first interface is of the individual feedback type and accessible only by the participant and contains the following functionalities:

A simplified data part with feedback with information refreshed every 30 seconds Fig. 4.
- Contract type
- Apparent and Active Power
- Daily and monthly consumption/production
- Daily and monthly cost
- Prediction of the day's consumption
- Applied pricing

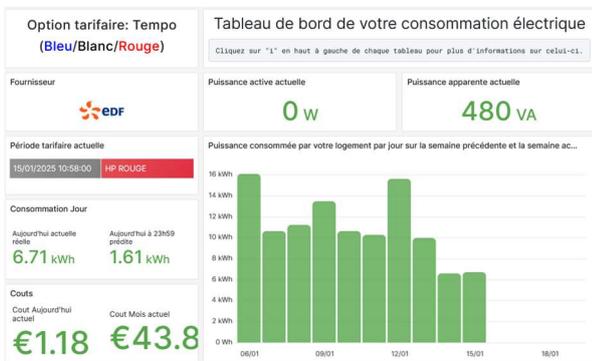

Fig. 4 User interface (simplified information part)

A historicization part allows for finding all the data generated by the experiment allowing them to make a posterior analysis of their behaviour Fig. 5.

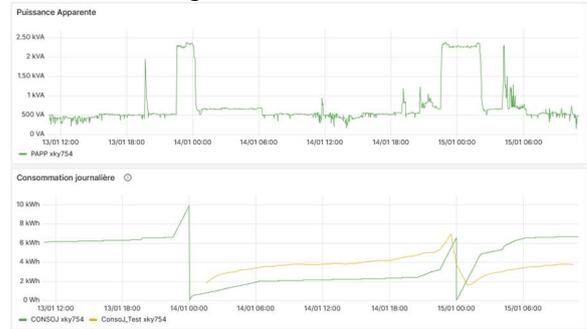

Fig. 5 User interface (historical information part)

The second interface is of the collective type and allows all participants to know the status of the operation. This interface is coupled with supervision allowing the diagnosis of the failure of one or more gateways. As with the individual interface, this is divided into 2 parts.

A first part with simplified and readable data where we find the following information Fig. 6:
- Overall current Apparent and Active Power
- Consumption/Production at different time horizon: daily, monthly and since the start of the operation
- Monitoring the system status

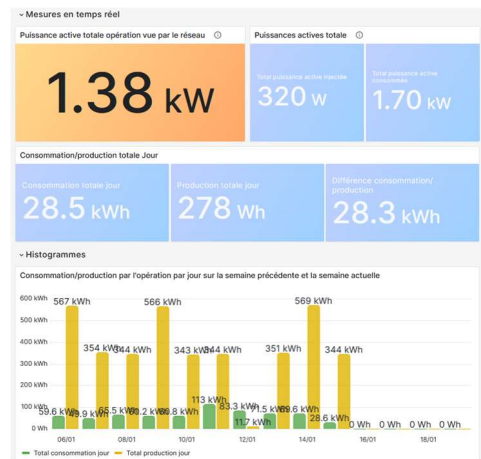

Fig. 6 Supervision interface (simplified information part)

A second part with an analysis of the power drawn or reinjected as well as the self-consumption and self-production rates of the operation Fig.7.

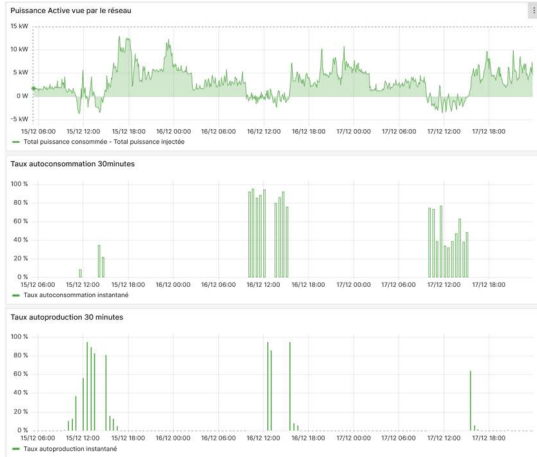

Fig. 7 Supervision interface (historical information part)

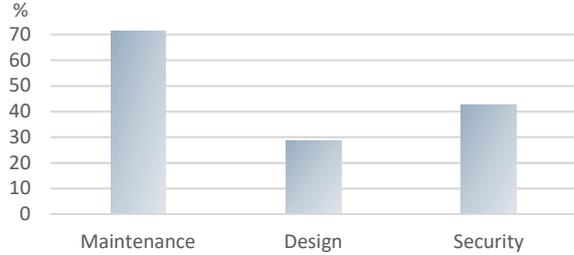

Fig. 8 Percentage of participants who asked a question in one of the 3 categories

The self-consumption and self-sufficiency rates are calculated on 3 different time scales (instantaneous, 30 min and display period) via the following calculation e.g. (1) and e.g. (2) based on the work of Jonathan Coignard [9].

$$self-consumption = \frac{\sum_{t=1}^{T} \min(load(t), prod(t))}{\sum_{t=1}^{T} prod(t)}$$

e.q. 1 Calculation of the self-consumption rate over a period

$$self-sufficiency = \frac{\sum_{t=1}^{T} \min(load(t), prod(t))}{\sum_{t=1}^{T} load(t)}$$

e.g. 2 Calculation of the self-sufficiency rate over a period

## 3 Results

*3.1 Feedback from the installation*

The first observation was that the disparity of knowledge (for example: some participants are very geeky and take a lot of initiatives while others did not know where their electricity meter is) brings a challenge in the creation of collective support in order to level the knowledge of each by providing relevant and useful information to the group. This support was also necessary for some participants during all stages of the process because the participants were afraid of doing wrong or of "having their data hacked".

During the implementation, 3 main groups of questions were raised by those accompanying the implementation of the device Fig. 8.
- Maintenance: Are there updates to be made? Do you have to reboot a sensor after a power outage or disconnection? Is there a manual backup of data to be made?

- Design: Is there an xKy equivalent on the market? Can everyone buy an xKy or equivalent? Do I have to pay extra to install this sensor?

- Security: Can I intercept my neighbor's data if I stand near the sensor? Can you (OTE, G2ELab researchers) turn off a device remotely with the xKy? Does installing the xKy sensor increase the risk of fire? Is installing this sensor legal and authorized by my network manager / Enedis?

The group of questions around security shows the reminiscences of the fears that already existed during the first deployments of the Linky meter [10] and shows the role of support is necessary to reassure the participants.

*3.2 Feedback on the use of interfaces*

In a previous experiment [11], we had started to test the different methods to incite participants to schedule their consumption. We started to applied this method on the first experiment xKy with the aim to encourage to connect to their interface and we had obtained an initial result where the type of newsletters sent to participants influenced the reconnection rate on the interfaces.
3 types of newsletters were sent:
- Newsletter Info: Announcing the status of the experiment
- Newsletter New Feature: Announcing the addition of a new feature
- Newsletter Action: Asking participants to do something on the interface

The following figure Fig.9 shows for the period March 2024/December 2024 that the first 2 types of newsletters (Info(b) and New Feature(c)) did not record a significant increase in the rate of reconnection to the interfaces unlike the newsletter Action (d).

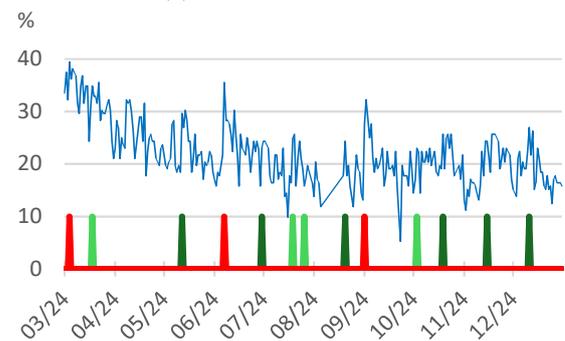

Fig. 9 (a) Percentage of participants logging into their interface at least once a day in blue, (b) newsletter Info in light green, (c) newsletter New Feature in dark green, (d) newsletter Action in red

For the collective self-consumption operation, we have not yet set up newsletters to have feedback on the rate of use of the interfaces. As it is shown in Fig. 10, for the period November 2024/January 2024 the daily connection rate is low because only the most technophile person in the group connects almost every day to his interface.

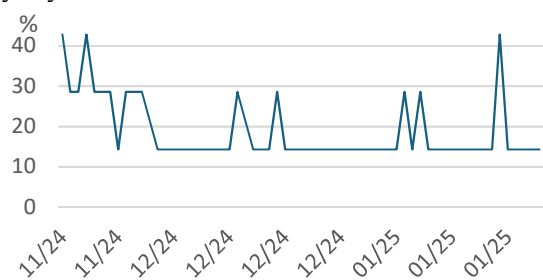

Fig. 10 Percentage of participants logging into their interface at least once a day

## 4 Conclusion

The xKy device allows real-time data from the collective self-consumption operation. It opens the way to indirect consumption management to optimize the self-consumption rate. A support is still needed to help people who do not have the knowledge to set it up. With this initial feedback, an improvement of the device is underway.
Currently, the rate of use of the system is low and shows that support or nudges may be necessary to complete it to move the system from observer status to operational flexibility actor.

## 5 Perspectives

The next step in the ongoing collective self-consumption experiment is to test the different alert channels to delay or even reduce consumption (email, SMS, notifications).
An accompanying document and a FAQ are being developed to enable simpler deployment of make the system trustworthy.

## 5 Acknowledgements

The xKy project is funded by Grenoble-INP SA and supported by UGA, CNRS and Observatory of Energy Transition ANR-15-IDEX-02. This project has received financial support from the CNRS through the MITI interdisciplinary programs through the project SITE-HIFI-CONS-OTE supported by the "Série de Longue Durée" program.